# Relaxed Local Correctability from Local Testing

Vinayak M. Kumar*      Geoffrey Mon†

**Abstract**

We construct the first asymptotically good relaxed locally correctable codes with polylogarithmic query complexity, bringing the upper bound polynomially close to the lower bound of Gur and Lachish (SICOMP 2021). Our result follows from showing that a high-rate locally testable code can boost the block length of a smaller relaxed locally correctable code, while preserving the correcting radius and incurring only a modest additive cost in rate and query complexity. We use the locally testable code's tester to check if the amount of corruption in the input is low; if so, we can "zoom-in" to a suitable substring of the input and recurse on the smaller code's local corrector. Hence, iterating this operation with a suitable family of locally testable codes due to Dinur, Evra, Livne, Lubotzky, and Mozes (STOC 2022) yields asymptotically good codes with relaxed local correctability, arbitrarily large block length, and polylogarithmic query complexity.

Our codes asymptotically inherit the rate and distance of any locally testable code used in the final invocation of the operation. Therefore, our framework also yields nonexplicit relaxed locally correctable codes with polylogarithmic query complexity that have rate and distance approaching the Gilbert–Varshamov bound.

## 1 Introduction

Locally correctable codes (LCCs) and locally decodable codes (LDCs) are error correcting codes that allow any bit of the original codeword or message, respectively, to be recovered using very few queries to a noisy codeword with bounded corruption. This is a natural and useful property, but unfortunately little is known about the best possible parameter tradeoffs between the rate and query complexity. In the asymptotically good (constant rate and distance) regime, Katz and Trevisan [KT00] show that any LDC (and any linear LCC) with block length $n$ must make $\tilde{\Omega}(\log n)$ queries. However, the most query-efficient constant-rate LCCs and LDCs, constructed by Kopparty, Meir, Ron-Zewi, and Saraf [KMRS17], require $2^{\tilde{O}(\sqrt{\log n})}$. Whether the true optimal query complexity is polylogarithmic or not is a longstanding open problem. A Reed–Muller code with appropriate parameters brings us tantalizingly close: such a code is locally correctable with polylogarithmic query complexity, but with block length slightly superlinear in the rate (see for e.g. [Yek12, Section 2.3]).

Ben-Sasson, Goldreich, Harsha, Sudan, and Vadhan [BGH+06] and Gur, Ramnarayan, and Rothblum [GRR20] introduced the notions of *relaxed* locally decodable codes (RLDCs) and *relaxed* locally correctable codes (RLCCs), respectively. These codes admit local decoders or correctors that either decode/correct, or detect corruption in the input by returning a rejection symbol ⊥. For asymptotically good RLDCs and (linear) RLCCs, the gap between lower and upper bounds is smaller but still significant: the best lower bound is $\tilde{\Omega}(\sqrt{\log n})$ due to Gur and Lachish [GL21], while the best upper bound, due to Cohen and Yankovitz [CY22], is $(\log n)^{O(\log \log \log n)}$. In this work, we improve the upper bound by constructing asymptotically good RLCCs with polylogarithmic query complexity.

**Theorem** (informal, see Corollary 4.1). *For infinitely many positive $n$ and any constant $R \in (0, 1)$, there exist explicit binary linear RLCCs (and thus RLDCs) of block length $n$, rate $R$, constant correcting (or decoding) radius, and query complexity $O(\log^{69} n)$.*

---

*Department of Computer Science, University of Texas at Austin. vmkumar@cs.utexas.edu.
†Department of Computer Science, University of Texas at Austin. gmon@cs.utexas.edu.

We make no effort to optimize the exponent, instead striving for a simpler exposition. The related and well-studied notion of locally testable codes (LTCs), where errors can be detected with few queries, proves to be key: we can build a relaxed local correctable code from any family of high-rate linear locally testable codes. Then, by leveraging the locally testable codes of Dinur, Evra, Livne, Lubotzky, and Mozes [DEL+22], we construct explicit RLCCs with constant rate, constant distance, and polylogarithmic query complexity. We also get nonexplicit RLCCs with polylogarithmic query complexity that approach the Gilbert–Varshamov bound, which is the best known general tradeoff between rate and distance for which codes exist. The last known RLCCs to approach the Gilbert–Varshamov bound are LCCs by Gopi, Kopparty, Oliveira, Ron-Zewi, and Saraf [GKO+18] which require polynomially many (i.e., $n^\varepsilon$) queries.

## 1.1 Techniques

To construct an asymptotically good RLCC with an arbitrarily large block length $n$, we start with a code with extremely small block length (say, polylogarithmic in $n$). Such a code is trivially an RLCC with polylogarithmic query complexity, because it has a corrector which reads all polylog $n$ symbols of the input. Then, we apply a transformation that uses this small RLCC to build an RLCC with slightly larger block length. We can repeat this transformation, building RLCCs with successively larger block length, until we reach our target block length $n$. At each of these steps, we need to maintain the distance and correcting radius of our code while minimizing the gain in query complexity and loss in rate, so that the final code will be asymptotically good. Indeed, such an approach has been used to construct asymptotically good LTCs and LCCs [Mei08, KMRS17], and is the basis for both prior asymptotically good RLCC constructions [GRR20, CY22].

### 1.1.1 Prior Constructions

Gur, Ramnarayan, and Rothblum [GRR20] use the tensor product in this framework to construct RLCCs. Starting from a RLCC with tiny block length, they build successively larger RLCCs by taking the tensor product of the code with itself. If $C \subseteq \mathbb{F}^n$ is a linear code, then the tensor product code $C \otimes C \subseteq \mathbb{F}^{n \times n}$ can be viewed as the set of matrices where every row and every column is a codeword of $C$. They show that if $C$ is an RLCC, $C \otimes C$ is also an RLCC because it has a local corrector that calls the corrector for $C$ as a subroutine on relevant rows and columns of the input; the corrector for $C \otimes C$ recurses on the smaller code's corrector polylog $n$ times. The tensor product step squares the block length each time, so $O(\log \log n)$ iterations are required to construct an RLCC with block length $n$; each step incurs a polylog $n$ factor in the query complexity, so the total queries required is $(\log n)^{O(\log \log n)}$.

Each tensor product step incurs a polylog $n$ factor in the query complexity because of its rate deterioration. If $C$ has rate $R = 1 - \varepsilon$, then $C \otimes C$ has rate $R^2 \approx 1 - 2\varepsilon$; the loss in rate has doubled after one iteration, so this loss will grow exponentially in the number of iterations. Therefore, for the final RLCC to have constant rate after $O(\log \log n)$ tensoring steps, the initial RLCC must have extremely high rate, at least $1 - 1/\log n$, which limits the distance of the initial RLCC. This initial distance causes the polylog $n$ multiplicative overhead in query complexity. To address this, Cohen and Yankovitz [CY22] give an alternative to tensoring called *row-evasive partitioning*, which incurs an *additive* loss in rate rather than a multiplicative one. The starting code can now have lower rate, like $1 - 1/\log \log n$, and thus higher distance, so the query overhead of each step (which is still a multiplicative factor) is improved to poly($\log \log n$). This yields a total query complexity of $(\log \log n)^{O(\log \log n)} = (\log n)^{O(\log \log \log n)}$.

### 1.1.2 Our Construction

To achieve asymptotically good RLCCs with polylog $n$ query complexity, we develop a new operation for boosting the block length of an RLCC. In contrast to tensoring and row-evasive partitioning, our operation augments the block length of an RLCC by a very modest polylog $n$ factor, but incurs only an *additive* cost in *both* rate and query complexity. Each step solely requires polylog $n$ additional queries, so although this operation needs to be iterated more often than tensoring, i.e., up to $m = O(\log n)$ times to reach block length $n$, the final query complexity is $m \cdot \text{polylog } n = \text{polylog } n$.

Intuitively, tensoring and row-evasive partitioning are operations that work by intertwining the structure of many copies of a smaller RLCC, so that the local corrector of the smaller RLCC can be used to cross-check overlapping portions of the input against each other. This necessitates multiple recursive calls to a smaller code's corrector at each step of the construction, which causes the final query complexity to grow exponentially in the number of iterations. Instead of cross-checking smaller RLCCs, our boosting operation enlists the outside help of a locally testable code to add structure. We will use the LTC's self-contained testing algorithm to ensure that it is safe to recurse on the smaller RLCC's corrector exactly once. This "tail recursion" is why our total query complexity grows linearly in the number of iterations, rather than exponentially.

Our boosting step is called *nesting*. Say we have an RLCC $C \subseteq \Sigma^n$ with correcting radius $\delta$ and a locally testable code $\text{LTC} \subseteq \Sigma^N$ with distance $2\delta$ where (for simplicity) $n$ divides $N$. Then, the code formed by *nesting $C$ in* LTC is defined to be

$$\text{LTC} \mathbin{\text{m}} C := \text{LTC} \cap C^{N/n},$$

and we claim that this code is an RLCC with correcting radius $\delta$. That is, this code has a local algorithm that, given any input $w$ that satisfies $\text{dist}(w, \text{LTC} \mathbin{\text{m}} C) \leq \delta$, either corrects or detects corruption.

To see why, suppose the distance from $w \in \Sigma^N$ to the nearest codeword $c \in \text{LTC} \mathbin{\text{m}} C$ is very small, i.e., $\text{dist}(w, \text{LTC} \mathbin{\text{m}} C) \leq \delta n/N$. Then we can hope to correct any index of $w$ by resorting to the local corrector of $C$. To correct $w_i$, we consider the unique interval $I := \{kn + 1, \ldots, kn + n\}$ containing $i$. We know by construction that $c|_I$ is a codeword of $C$. Furthermore, we know that $w$ differs from $c$ in at most $\delta n$ indices, so $\text{dist}(w|_I, c|_I) \leq \delta$ which is within the correcting radius of $C$. Hence, we can "zoom-in" and use the corrector for $C$ to correct any symbol of $w|_I$, including $w_i$.

Otherwise, if $\delta n/N < \text{dist}(w, \text{LTC} \mathbin{\text{m}} C) \leq \delta$, we can hope to detect (rather than correct) corruption using the local tester of LTC. A locally testable code, by definition, has a local testing algorithm $T$ that rejects its input $w$ with probability proportional to $\text{dist}(w, \text{LTC})$. Because $\text{LTC} \mathbin{\text{m}} C \subseteq \text{LTC}$ and LTC has distance $2\delta$, we know that $\text{dist}(w, \text{LTC}) = \text{dist}(w, \text{LTC} \mathbin{\text{m}} C) > \delta n/N$, and so $T$ will reject $w$ with probability $\Omega(\delta n/N)$. Thus, $O(N/\delta n)$ repetitions (hiding some minor factors) of the local tester suffice to detect corruption with constant probability.

Therefore, by combining the corrector for $C$ and tester for LTC, we can handle both cases. Run both the corrector and the tester; if the tester finds corruption, we can return the reject symbol $\perp$, and otherwise we are likely in the small distance case and can return the output of the corrector. This shows that $\text{LTC} \mathbin{\text{m}} C$ is an RLCC which inherits the larger block length of LTC while requiring only roughly $O(N/\delta n)$ more queries than $C$.

Furthermore, we can show that nesting does not destroy the rate. If LTC and $C$ are linear, and if LTC has rate $1 - \varepsilon_{\text{LTC}}$ while $C$ has rate $1 - \varepsilon$, then by counting the number of linear constraints, $\text{LTC} \mathbin{\text{m}} C$ has rate at least $1 - \varepsilon - \varepsilon_{\text{LTC}}$. If we hope to repeatedly apply nesting with larger and larger LTCs (block length growing by a small factor) to build an RLCC with arbitrarily large block length $n$, then we will need to apply nesting $m = O(\log n)$ times. Assuming that all of the LTCs have rate $1 - \varepsilon_{\text{LTC}}$, the RLCC will have rate of $1 - O(\log n) \cdot \varepsilon_{\text{LTC}}$ which we need to be $\Omega(1)$ in order to be asymptotically good. This forces us to use LTCs with rate at least $1 - O(1/\log n)$.

Fortunately, we can use the construction of Dinur, Evra, Livne, Lubotzky, and Mozes [DEL$^+$22] to get the high-rate linear LTCs we need. For any sufficiently large choice of $n$, there is an infinite sequence of explicit LTCs with rate at least $1 - O(1/\log n)$ and with appropriate local testing parameters so that the additive query overhead of each nesting step is polylog $n$ as desired. By iteratively nesting these LTCs, we can build an RLCC of block length $n$ with constant rate and polylogarithmic query complexity.

One may notice that since the LTCs have rate $1 - o(1)$, they must also have $o(1)$ distance, because of e.g. the Singleton bound. The correcting radius of our RLCC thus will also be $o(1)$, and so our code is not yet asymptotically good. This is remedied by nesting in one last LTC with constant rate and distance. This boosts the correcting radius to be constant, and the rate of this last LTC dominates the rate of the final RLCC, which at last is asymptotically good.

## 1.2 Related Work

**Prior constructions and lower bounds.** The two main parameter regimes for RLDCs and RLCCs are the constant query regime (optimizing block length given the dimension $k$ and query complexity $q$) and the asymptotically good regime (optimizing query complexity when $k/n = \Theta(1)$ and $\delta = \Theta(1)$). In the constant-query regime, the best known block length is $n = O(k^{1+O(1/q)})$ [AS21], following a line of work [BGH+06, GRR20, CGS22]. Interestingly, this matches (up to a constant factor in $q$) the block length lower bound for full-fledged LDCs [KT00, Woo07].

Table 1 summarizes the historic state of the art query complexity for asymptotically good RLCCs. This table does not include Gopi, Kopparty, Oliveira, Ron-Zewi, and Saraf [GKO+18] who optimize rate (approaching the Gilbert–Varshamov bound) instead of queries.

Gur and Lachish [GL21] establish lower bounds for RLDCs, including a $\tilde{\Omega}(\sqrt{\log n})$ query lower bound for asymptotically good RLDCs with nonadaptivity. Dall'Agnol, Gur, and Lachish [DGL21] extend the lower bound to adaptive decoders, and Goldreich [Gol23b] provides an alternative and simpler proof, which in some cases is stronger. Block, Blocki, Cheng, Grigorescu, Li, Zheng, and Zhu [BBC+23] prove an exponential block length lower bound for 2-query RLDCs, asymptotically matching the exponential block length lower bound for 2-query LDCs established by Kerenidis and de Wolf [KW03].

**Alternative error models.** LDCs, LCCs, and their relaxed counterparts have been studied in other error models, distinct from the Hamming worst-case error setting that we study in this work. These codes have been studied in the insertion-deletion error model, where a limited number of symbols can be added or removed (rather than simply flipped) anywhere in the codeword [OP15, BBG+20, CLZ20, BBC+23]. In addition, both the Hamming and insertion-deletion models have been studied in the computationally bounded setting, where the adversary choosing the location of bit flips or insertions/deletions has limited resources. Then, cryptographic assumptions can be used to construct LDCs and LCCs [OPS07, HO08, HOSW11, BKZ20, BB21, ABB22] as well as their relaxed counterparts [BGGZ21, BB23].

In particular, the latter two works use these assumptions to (among other results) construct asymptotically good RLDCs and RLCCs in the computationally bounded Hamming model with polylogarithmic queries. We achieve this in the general setting.

**Subsequent work.** Soon after we posted the preprint for this work, Cohen and Yankovitz [CY23] improved our technique and lowered the query complexity for asymptotically good RLCCs from our $\log^{69} n$ to $(\log n)^{2+o(1)}$. They observe that the nesting operation can boost an RLCC using a code that is locally testable only on inputs with bounded corruption, rather than a full-fledged locally testable code which must work on all inputs. Thus, they can use a family of expander codes, which satisfy this weaker property, to simplify the RLCC construction and improve the query complexity.

| Technique | Query complexity | Due to |
|---|---|---|
| low-degree polynomials | $n^\varepsilon$ | [BFLS91, RS96] |
| multiplicity codes | $n^\varepsilon$ | [KSY14] |
| lifted Reed–Solomon codes | $n^\varepsilon$ | [GKS13] |
| expander graphs | $n^\varepsilon$ | [HOW15] |
| distance amplification | $2^{\tilde{O}(\sqrt{\log n})}$ | [KMRS17] |
| repeated tensoring | $(\log n)^{O(\log \log n)}$ | [GRR20] |
| row-evasive partitions | $(\log n)^{O(\log \log \log n)}$ | [CY22] |
| nested LTCs | $\log^{O(1)} n$ | this work |

Table 1: Best known query complexity for asymptotically good RLCCs with block length $n$.

# 2 Preliminaries

## 2.1 General Notation

Let dist($x, y$) denote the relative Hamming distance between two strings with the same length and alphabet:

$$\forall x, y \in \Sigma^n. \ \mathrm{dist}(x, y) := \frac{\#\{i \in [n] : x_i \neq y_i\}}{n}.$$

All of the necessary properties of a distance function are satisfied, including the triangle inequality. For every subset $S \subseteq \Sigma^n$, let $\mathrm{dist}(x, S) := \min_{y \in S} \mathrm{dist}(x, y)$.

We say that $f(n) \leq \mathrm{poly}(n)$ if there is a fixed polynomial $p$ such that for large enough $n$, $f(n) \leq p(n)$, and analogously for $\geq$. We say $f(n) = \mathrm{poly}(n)$ if $f(n) \leq \mathrm{poly}(n)$ and $f(n) \geq \mathrm{poly}(n)$. Analogous conventions are used for polylog $n$, which denotes $\mathrm{poly}(\log n)$. The polynomials implicitly defined by poly or polylog are fixed with respect to all parameters involved.

Let $\mathbb{N}$ denote the set of positive integers, and $\mathbb{F}_2$ denote the finite field of 2 elements. For every $k \in \mathbb{N}$, define $[k] := \{1, 2, \ldots, k\}$. For a string $x \in \Sigma^n$ and an index set $I \subseteq [n]$, let $x|_I$ denote the restriction of $x$ to the indices in $I$. For a set of strings $S \subseteq \Sigma^n$, let $S|_I$ denote the set $\{x|_I : x \in S\}$.

## 2.2 Error-Correcting Codes

An *error-correcting code* is a set of strings, called *codewords*, where every two codewords are well-separated.

**Definition 2.1.** A code over an alphabet $\Sigma$ with dimension $k$, block length $n$, and distance $\delta$ is a subset $C \subseteq \Sigma^n$ of size $|\Sigma^k|$ where

$$\forall c \in C. \ \mathrm{dist}(c, C \setminus c) \geq \delta.$$

If $\Sigma = \{0, 1\}$, $C$ is called a *binary code*. In this work, we treat $\{0, 1\}$ and $\mathbb{F}_2$ as interchangeable.

The ratio $k/n$ is called the *rate* of a code. In this work, we study *asymptotically good* codes which have constant rate and distance.

*Remark* 2.2. If $C \subseteq \Sigma^n$ is a code with distance $\delta$, $w \in \Sigma^n$, and $c \in C$ satisfies $\mathrm{dist}(w, c) < \delta/2$, then $c$ must be the *unique* closest codeword to $w$. This is because for any other codeword $c' \in C$,

$$\mathrm{dist}(c', w) \geq \mathrm{dist}(c', c) - \mathrm{dist}(w, c) \geq \delta - \delta/2 > \mathrm{dist}(c, w)$$

by the triangle inequality.

**Definition 2.3.** A code $C \subseteq \Sigma^n$ is *systematic* if there is a bijective map $\mathrm{Enc} : \Sigma^k \to C$ where the message appears in the codeword:

$$\forall m \in \Sigma^k. \ \mathrm{Enc}(m)|_{[k]} = m.$$

**Definition 2.4.** A *linear code* is a code $C \subseteq \mathbb{F}^n$ which is a linear subspace over a finite field. Every linear code can be specified by a *generator matrix* which gives a basis for the code and which also serves as an encoding map, or by a *parity-check matrix* which gives a set of linear constraints that every codeword must satisfy. We say that a family of linear codes is *explicit* if each code's parity-check matrix can be computed uniformly in time polynomial to the size of the matrix.

*Remark* 2.5. A linear code can always be made systematic, because we can take its generator matrix, apply Gaussian elimination, and permute columns until it contains a $k \times k$ identity matrix in the first $k$ columns.

## 2.3 Relaxed Locally Correctable Codes

LCCs and LDCs can recover any bit of the closest codeword or message, respectively, using very few queries to a noisy codeword. The study of LCCs and LDCs has roots in program checking [BK95, Lip90, BFLS91, RS96] and was first formalized by Katz and Trevisan [KT00]; we refer the reader to Yekhanin's comprehensive survey [Yek12].

*Relaxed* locally decodable codes and *relaxed* locally correctable codes have the option of detecting corruption instead of decoding or correcting. The study of these codes is closely related to probabilistically checkable proofs and originates with Ben-Sasson, Goldreich, Harsha, Sudan, and Vadhan [BGH+06], who defined RLDCs and gave the first constructions.

**Definition 2.6** (RLDC). A code $C \subseteq \Sigma^n$ is a *relaxed locally decodable code* with decoding radius $\delta$ and query complexity $q$ if it has an encoding map $\text{Enc} : \Sigma^k \to C$ and a randomized corrector $M$ that makes $q$ queries such that

1. (Completeness) For every $m \in \Sigma^k$,

$$\forall i \in [k]. \ \Pr[M^{\text{Enc}(m)}(i) = m_i] = 1.$$

2. (Soundness) For every $m \in \Sigma^k$ and every $w \in \Sigma^n$ with $\text{dist}(w, \text{Enc}(m)) \leq \delta$,

$$\forall i \in [n]. \ \Pr[M^w(i) \in \{m_i, \bot\}] \geq \frac{2}{3}.$$

The superscript denotes the input that $M$ queries.

Gur, Ramnarayan, and Rothblum [GRR20] later introduced the corresponding notion of RLCCs along with constructions.

**Definition 2.7** (RLCC). A code $C \subseteq \Sigma^n$ is a *relaxed locally correctable code* with correcting radius $\delta$ and query complexity $q$ if it has a randomized corrector $M$ that makes $q$ queries such that

1. (Completeness) For every $c \in C$,

$$\forall i \in [n]. \ \Pr[M^c(i) = c_i] = 1.$$

2. (Soundness) For every $c \in C$ and $w \in \Sigma^n$ with $\text{dist}(w, c) \leq \delta$,

$$\forall i \in [n]. \ \Pr[M^w(i) \in \{c_i, \bot\}] \geq \frac{2}{3}.$$

The superscript denotes the input that $M$ queries. We refer to $M$ as a *C-corrector*.

Here, we have given strong definitions of RLDCs and RLCCs featuring perfect completeness. Goldberg [Gol23a] shows that for linear RLDCs and RLCCs, the above definition is essentially equivalent to allowing imperfect completeness (the corrector/decoder can sometimes err on true codewords) and requiring nonadaptivity (the corrector/decoder's queries do not depend on the outcome of prior queries). That said, all of the codes that we construct have nonadaptive correctors.

A systematic RLCC implies an RLDC with the same radius and query complexity, because we can use the local corrector on the portion of the codeword which corresponds to message symbols. In addition, an RLCC with correcting radius $\delta$ must have distance at least $\delta$ in order for perfect completeness and soundness to simultaneously hold. Therefore, we say an RLCC is asymptotically good if it has constant rate and constant correcting radius, which implies constant distance.

## 2.4 Locally Testable Codes

Locally testable codes (LTCs) are codes with testing algorithms that can gauge corruption with few queries to the input. We will make use of the following (strong) definition of LTCs:

**Definition 2.8** (LTC). A code $C \subseteq \Sigma^n$ is a *locally testable code* with distance $\delta$, testability[1] $\kappa$, and query complexity $q$ if it has a randomized tester $T$ that makes $q$ queries and returns either $\top$ (accept) or $\bot$ (reject), such that

1. (Completeness) For every $c \in C$,
$$\Pr[T^c = \top] = 1.$$

2. (Soundness) For every $w \in \Sigma^n$,
$$\Pr[T^w = \bot] \geq \kappa \cdot \mathrm{dist}(w, C).$$

The superscript denotes the input that $T$ queries. We refer to $T$ as a *C-tester*.

Dinur, Evra, Livne, Lubotzky, and Mozes [DEL+22] and Panteleev and Kalachev [PK22] constructed the first locally testable codes with constant rate, distance, and query complexity (referred to as $c^3$-LTCs). In particular, the former are able to construct explicit families of linear LTCs with rate arbitrarily close to 1:

**Theorem 2.9** ([DEL+22, Theorem 1.1 and Remark 5.3]). *For any $R = 1 - \varepsilon \in (0, 1)$, there is a prime power $p = \Theta((1/\varepsilon)^{10})$ and values $\delta \geq \Omega(\varepsilon^3)$, $\kappa \geq \Omega(\varepsilon^{15})$, and $q \leq O((1/\varepsilon)^{20})$, such that for all $j \in \mathbb{N}$, there exists a binary linear LTC $\mathsf{LTC}_j$ with rate at least $R$, minimum distance at least $\delta$, testability at least $\kappa$, query complexity at most $q$, and block length $n_j := (q/8) \cdot (p^{3j} - p^j)$.*

## 3 Construction

### 3.1 Boosting RLCC Block Length

The building block of our construction is the *nesting* operation ⋒ which combines two codes.

**Definition 3.1.** Let $C_1 \subseteq \Sigma^N$ and $C_2 \subseteq \Sigma^n$ with $n \leq N$. Define
$$C_1 \Cap C_2 := C_1 \cap (C_2^{\lfloor N/n \rfloor} \times \Sigma^{N - \lfloor N/n \rfloor \cdot n}) \cap (\Sigma^{N-n} \times C_2).$$
We say that $C_1 \Cap C_2$ is the code formed by *nesting $C_2$ in $C_1$*.

That is, $C_1 \Cap C_2$ is the subset of codewords $c$ in $C_1$ such that $c|_{\{1,\ldots,n\}} \in C_2$, $c|_{\{n+1,\ldots,2n\}} \in C_2$, and so on. If $n$ does not divide $N$, then we also require $c|_{\{N-n+1,\ldots,N\}} \in C_2$. Because $C_1 \Cap C_2$ is a subset of $C_1$, it must have distance at least that of $C_1$. In addition, when both codes are linear, we can bound the rate of $C_1 \Cap C_2$. Nesting $C_2$ in $C_1$ incurs an additive loss in rate:

**Lemma 3.2.** *Let $n \leq N$. Let $C_1 \subseteq \Sigma^N$ be a linear code with rate $1 - \varepsilon_1$ and distance $\delta_1$, and let $C_2 \subseteq \Sigma^n$ be a linear code with rate $1 - \varepsilon_2$. Then, $C_1 \Cap C_2 \subseteq \Sigma^N$ is a linear code with rate at least $1 - \varepsilon_1 - (n/N \cdot \lceil N/n \rceil)\varepsilon_2$.*

*Proof.* Using the rank-nullity theorem, the total number of linear constraints defining $C_1 \Cap C_2$ is
$$\leq \varepsilon_1 N + \varepsilon_2 n \cdot \left\lceil \frac{N}{n} \right\rceil = \left(\varepsilon_1 + \frac{n}{N} \cdot \left\lceil \frac{N}{n} \right\rceil \cdot \varepsilon_2\right) \cdot N$$
which then implies that the rate of $C_1 \Cap C_2$ is
$$\geq 1 - \varepsilon_1 - \frac{n}{N} \cdot \left\lceil \frac{N}{n} \right\rceil \cdot \varepsilon_2. \qquad \square$$

*Remark* 3.3. If $n$ divides $N$, then $C_1 \Cap C_2 = C_1 \cap C_2^{N/n}$, with rate at least $1 - \varepsilon_1 - \varepsilon_2$. When $n$ does not divide $N$, the rate of $C_1 \Cap C_2$ is at least $1 - \varepsilon_1 - (1 + n/N)\varepsilon_2$.

Next, we prove that nesting preserves relaxed local correctability when performed with an LTC. We can lift a smaller RLCC to a larger block length by nesting it inside an LTC.

---

[1]This parameter has also been referred to as the *detection probability* e.g. [DEL+22].

**Lemma 3.4.** *Let* LTC $\subseteq \Sigma^{n_{\text{LTC}}}$ *be an LTC with query complexity $q_{\text{LTC}}$, distance $\delta_{\text{LTC}}$, and testability $\kappa$. Let $C \subseteq \Sigma^n$ be an RLCC with query complexity $q$ and correcting radius $\delta$ where $n \leq n_{\text{LTC}}$. Then, LTC $\pitchfork C \subseteq \Sigma^{n_{\text{LTC}}}$ is an RLCC with correcting radius $\delta_{\text{LTC}}/2$ and query complexity $q + O(q_{\text{LTC}} n_{\text{LTC}}/\delta \kappa n)$.*

*Proof.* Given the LTC-tester $T$ and the $C$-corrector $M_C$, we can give a corrector $M^w(i)$ for LTC $\pitchfork C$:

1. Let $I$ be an interval of $[n_{\text{LTC}}]$ of size $n$, defined as follows:

$$I := \begin{cases} \{\lceil i/n \rceil \cdot n - (n-1), \ldots, \lceil i/n \rceil \cdot n\} & \text{if } i < n_{\text{LTC}} - n \\ \{n_{\text{LTC}} - (n-1), \ldots, n_{\text{LTC}}\} & \text{if } i \geq n_{\text{LTC}} - n \end{cases}$$

   This is chosen so that that $i \in I$ and $\forall c \in$ LTC $\pitchfork C$. $c|_I \in C$.

2. Run $M_C^{w|_I}(i^*)$, where $i^* := i + 1 - \min I$. We choose $i^*$ such that $(w|_I)_{i^*} = w_i$.

3. For $t := O(n_{\text{LTC}}/\delta \kappa n)$ iterations, run $T^w$.

4. Output $\bot$ if any run of the LTC-tester in step 3 returns $\bot$; otherwise output the result of step 2.

The query complexity of step 2 is $q$, while the query complexity of step 3 is $q_{\text{LTC}} \cdot O(n_{\text{LTC}}/\delta \kappa n)$. In addition, $M$ is nonadaptive as long as $T$ and $M_C$ are nonadaptive.

Next, we can show that $M$ satisfies the perfect completeness condition of the RLCC definition. Since $w \in$ LTC $\pitchfork C \subseteq$ LTC, every repetition of the LTC-tester in step 3 will return $\top$ (accept), so for all $i$, $M^w(i)$ will return the result of step 2. By definition, if $w \in$ LTC $\pitchfork C$, then $w|_I \in C$. Therefore, the $C$-corrector in step 2 will return $(w|_I)_{i^*} = w_i$ with certainty, and so will $M^w(i)$, as needed.

Finally, we show soundness. Let $0 < \text{dist}(w, \text{LTC} \pitchfork C) < \delta_{\text{LTC}}/2$, and let $c \in$ LTC $\pitchfork C$ be the unique closest codeword to $w$. Note that $c \in$ LTC and $c|_I \in C$. Then, for all $i$, we need to show $\Pr[M^w(i) \in \{c_i, \bot\}] \geq 2/3$.

- First, assume $\text{dist}(w, c) = \text{dist}(w, \text{LTC} \pitchfork C) \geq \delta n/2n_{\text{LTC}}$. Because $\text{dist}(w, c) < \delta_{\text{LTC}}/2$, $c$ is the unique codeword in LTC which is closest to $w$ by Remark 2.2. Hence, one run of $T^w$ will return $\bot$ with probability at least $\kappa \delta n/n_{\text{LTC}}$. Therefore, step 3 returns $\bot$ with probability

$$\geq 1 - \left(1 - \frac{\delta \kappa n}{2 n_{\text{LTC}}}\right)^t \geq 1 - \exp\left(-\frac{t \delta \kappa n}{2 n_{\text{LTC}}}\right) \geq 2/3,$$

  for a suitable constant in $t$. Thus, $\forall i. \Pr[M^w(i) = \bot] \geq 2/3$ which satisfies soundness.

- Now assume $\text{dist}(w, c) < \delta n/2 n_{\text{LTC}}$. From here we can bound the distance between the substrings $w|_I$ and $c|_I$; the distance of the substrings is maximized if all of the symbols that differ between $w$ and $c$ lie in the interval $I$. Consequently, $\text{dist}(w|_I, c|_I) < \delta/2$, and $c|_I$ is the unique codeword in $C$ closest to $w|_I$ (see Remark 2.2).

  Applying the soundness condition of the $C$-corrector,

$$\Pr[M_C^{w|_I}(i^*) \in \{\underbrace{(c|_I)_{i^*}}_{c_i}, \bot\}] \geq 2/3.$$

  $M^w(i)$ will return either the output of $M_C^{w|_I}(i^*)$, or $\bot$ if some iteration of step 3 rejects. Hence,

$$\Pr[M^w(i) \in \{c_i, \bot\}] \geq \Pr[M_C^{w|_I}(i^*) \in \{c_i, \bot\}] \geq 2/3.$$

This shows that the soundness condition is satisfied in all cases. $\square$

In summary, nesting an RLCC in an LTC yields a code which inherits the best properties of both: the resulting code inherits distance and block length from the larger LTC, as well as the relaxed local correctability of the smaller RLCC. Therefore, we can build an RLCC with arbitrarily large block length by iteratively nesting the code in a series of LTCs with larger and larger block length.

**Proposition 3.5.** *Let* $\text{LTC}_1, \ldots, \text{LTC}_m$ *be a sequence of linear LTCs over a finite field alphabet $\mathbb{F}$ with block lengths $n_1, \ldots, n_m$, which satisfy the following properties:*

- $n_j \leq n_{j+1}$ for all $j$,
- every $\mathsf{LTC}_j$ has rate at least $1 - \varepsilon_{\mathsf{LTC}}$ and distance at least $\delta_{\mathsf{LTC}}$, and
- every $\mathsf{LTC}_j$ has a local tester with query complexity at most $q_{\mathsf{LTC}}$ and testability at least $\kappa_{\mathsf{LTC}}$.

Then, $C := \mathsf{LTC}_m \mathbin{\text{\m@}} (\mathsf{LTC}_{m-1} \mathbin{\text{\m@}} \cdots (\mathsf{LTC}_2 \mathbin{\text{\m@}} \mathsf{LTC}_1)\dots)$ is a linear RLCC which satisfies the following properties:

- alphabet $\mathbb{F}$ and block length $n_m$,
- rate at least

$$1 - \varepsilon_{\mathsf{LTC}}\left(1 + \sum_{j=2}^{m} \prod_{j'=2}^{j} \frac{n_{j'-1}}{n_{j'}} \left\lceil \frac{n_{j'}}{n_{j'-1}} \right\rceil \right),$$

- correcting radius at least $\delta_{\mathsf{LTC}}/2$, and
- query complexity at most

$$n_1 + \sum_{j=2}^{m} O\!\left(\frac{q_{\mathsf{LTC}} n_j}{\delta_{\mathsf{LTC}} \kappa_{\mathsf{LTC}} n_{j-1}}\right).$$

*Remark* 3.6. If $n_j$ divides $n_{j+1}$ for all $j$, then the rate lower bound simplifies to $1 - m\varepsilon_{\mathsf{LTC}}$. Even when $n_j$ does not divide $n_{j+1}$, with appropriate parameter choices we can still simplify the rate to $1 - O(m\varepsilon_{\mathsf{LTC}})$. Therefore, we can view each nesting step as reducing the rate of $C$ by $O(\varepsilon_{\mathsf{LTC}})$, so that each step incurs an additive (rather than multiplicative) cost in both rate and query complexity.

*Proof.* The alphabet and block length of $C$ follows from the definition of nesting. Next, we show the rate. Let $\mathsf{LTC}_j \mathbin{\text{\m@}} \cdots (\mathsf{LTC}_2 \mathbin{\text{\m@}} \mathsf{LTC}_1)$ have rate $1 - \varepsilon_j$. We know $\varepsilon_1 \leq \varepsilon_{\mathsf{LTC}}$ by definition, and Lemma 3.2 tells us that for $j \geq 2$,

$$\varepsilon_j \leq \varepsilon_{\mathsf{LTC}} + \frac{n_{j-1}}{n_j}\left\lceil \frac{n_j}{n_{j-1}} \right\rceil \cdot \varepsilon_{j-1}.$$

Hence by induction, $\varepsilon_m$ is upper bounded by the series

$$\varepsilon_m \leq \varepsilon_{\mathsf{LTC}} \cdot \left(1 + \sum_{j=2}^{m} \prod_{j'=2}^{j} \frac{n_{j'-1}}{n_{j'}} \left\lceil \frac{n_{j'}}{n_{j'-1}} \right\rceil \right),$$

which gives the desired lower bound on the rate of $C$.

Finally, we show that $C$ is an RLCC. $\mathsf{LTC}_1$ can be viewed as an RLCC with correcting radius $\delta_{\mathsf{LTC}}$ because it has a trivial correcting algorithm that reads the entire input and checks whether it is in $\mathsf{LTC}_1$. This corrector has query complexity $n_1$. Then by applying Lemma 3.4, $\mathsf{LTC}_2 \mathbin{\text{\m@}} \mathsf{LTC}_1$ has a corrector with correcting radius $\delta_{\mathsf{LTC}}/2$ and query complexity

$$\leq n_1 + O\!\left(\frac{q_{\mathsf{LTC}} n_2}{\delta_{\mathsf{LTC}} \kappa_{\mathsf{LTC}} n_1}\right).$$

By applying this lemma $m - 2$ more times with $\mathsf{LTC}_3, \dots, \mathsf{LTC}_m$, we can conclude that there is a $C$-corrector with correcting radius $\delta_{\mathsf{LTC}}/2$ and total query complexity

$$\leq n_1 + \sum_{j=2}^{m} O\!\left(\frac{q_{\mathsf{LTC}} n_j}{\delta_{\mathsf{LTC}} \kappa_{\mathsf{LTC}} n_{j-1}}\right). \qquad \square$$

## 3.2 Instantiating the Nesting Framework

We now have all of the tools we need to construct asymptotically good RLCCs, as long as we pick a suitable sequence of LTCs. If the block length of each successive LTC grows by at least a constant factor, then $m \leq O(\log n)$. Thus, using Remark 3.6 as guidance, we need a family of explicit LTCs with rate at least $1 - O(1/\log n)$ in order for Proposition 3.5 to yield an RLCC with constant rate. This means these LTCs will have subconstant distance, due to the Singleton bound. We will use the following instantiation of Theorem 2.9:

**Corollary 3.7** ([DEL+22], see Appendix A). *For every sufficiently large $N \in \mathbb{N}$, there exists an integer $n \in [\Omega(N/\log^{30} N), N]$ and a family of LTCs $\{\mathsf{LTC}_1, \ldots, \mathsf{LTC}_m\}$ such that*

- *Each $\mathsf{LTC}_j$ is binary, linear, and explicit.*
- *Each $\mathsf{LTC}_j$ has:*
  - *rate $R = 1 - \varepsilon_{\mathsf{LTC}} \geq 1 - O(1/\log N)$,*
  - *distance $\delta_{\mathsf{LTC}} \geq \Omega(1/\log^3 N)$,*
  - *testability $\kappa_{\mathsf{LTC}} \geq \Omega(1/\log^{15} N)$, and*
  - *query complexity $q_{\mathsf{LTC}} \leq O(\log^{20} N)$.*
- *Each $\mathsf{LTC}_j$ has block length $n_j$, such that $n_1 \leq O(\log^{50} N)$, $n_m = n$, and $\forall j. \ n_{j+1}/n_j = \Theta(\log^{30} N)$.*
- *The number of codes is $m = O(\log N / \log \log N)$.*

*Remark* 3.8. In this concrete family of LTCs, the block length of each successive LTC increases by a polylogarithmic factor, instead of the constant factor we previously imagined using. Thus, $m = O(\log n / \log \log n) = o(\log n)$. The $n_j / n_{j-1} = \mathrm{polylog}\, n$ factor also appears in the query complexity, which is acceptable. One could choose the LTC rate to be $1 - O(\log \log n / \log n)$ instead in order to marginally improve parameters such as $\delta_{\mathsf{LTC}}, \kappa_{\mathsf{LTC}}$, and $q_{\mathsf{LTC}}$ which appear in the query complexity, while worsening the rate.

With this family of LTCs, we can construct high rate RLCCs with polylogarithmic query complexity, albeit with subconstant correcting radius.

**Theorem 3.9.** *For sufficiently large $N \in \mathbb{N}$, there is an explicit binary linear RLCC with block length $n \in [\Omega(N/\log^{30} N), N]$, rate $1 - O(1/\log \log N)$, correcting radius $\Omega(1/\log^3 N)$, and query complexity $O(\log^{69} N / \log \log N)$.*

*Proof.* We can use the parameter $N$ with Corollary 3.7 to get a family of LTCs to plug into Proposition 3.5. Then, our code is $C \coloneqq \mathsf{LTC}_m \pitchfork (\mathsf{LTC}_{m-1} \pitchfork \ldots (\mathsf{LTC}_2 \pitchfork \mathsf{LTC}_1) \ldots )$, with parameters as defined in Corollary 3.7. We find that our code $C$ has block length $n \in [\Omega(N/\log^{30} N), N]$ and rate at least

$$\geq 1 - \varepsilon_{\mathsf{LTC}} \left(1 + \sum_{j=2}^{m} \prod_{j'=2}^{j} \left(\frac{n_{j'-1}}{n_{j'}} \left\lceil \frac{n_{j'}}{n_{j'-1}} \right\rceil \right)\right)$$

$$\geq 1 - \varepsilon_{\mathsf{LTC}} \left(1 + \sum_{j=2}^{m} \prod_{j'=2}^{j} \left(1 + \frac{n_{j'-1}}{n_{j'}}\right)\right)$$

$$\geq 1 - \varepsilon_{\mathsf{LTC}} \sum_{j=1}^{m} \left(1 + O(1/\log^{30} N)\right)^{j-1}$$

$$\geq 1 - \varepsilon_{\mathsf{LTC}} m \cdot \exp\left(O\left(\frac{m}{\log^{30} N}\right)\right)$$

$$= 1 - (1 + o(1)) m \varepsilon_{\mathsf{LTC}}$$

$$= 1 - O\left(\frac{1}{\log \log N}\right).$$

In addition, $C$ is an RLCC with radius $\delta_{\mathsf{LTC}}/2 = \Omega(1/\log^3 N)$ and query complexity

$$\leq n_1 + \sum_{i=2}^{m} O\left(\frac{q_{\mathsf{LTC}} n_i}{\delta_{\mathsf{LTC}} \kappa_{\mathsf{LTC}} n_{i-1}}\right)$$

$$\leq O(\log^{50} N) + \sum_{i=2}^{m} O\left(\log^{20+30+3+15} N\right)$$

$$\leq O\left(\frac{\log^{69} N}{\log \log N}\right). \qquad \square$$

$C$ is an RLCC with the desired rate and query complexity, but it is not yet asymptotically good since its correcting radius is determined by the distance of the LTC family from Corollary 3.7. This distance is subconstant since all of the LTCs must have rate $1 - O(1/m)$. Thankfully, we can boost the correcting radius of our RLCC with one final nesting step using an LTC with constant distance and rate. The increase in query complexity is negligible, and the rate of the last LTC dominates the rate of the resulting RLCC.

**Corollary 3.10.** *Let* LTC *be a binary linear LTC with sufficiently large block length $N$, and with rate $R$, distance $\delta_{\text{LTC}}$, testability $\kappa$, and query complexity $q$. Then, there exists a binary linear RLCC $C$ with block length $N$, rate $R - O(1/\log \log N)$, correcting radius $\delta_{\text{LTC}}/2$, and query complexity*

$$O\left(\frac{q}{\kappa} \cdot \log^{33} N + \frac{\log^{69} N}{\log \log N}\right).$$

*$C$ is explicit if and only if* LTC *is explicit.*

*Proof.* We pick sufficiently large $N$ such that we can use Theorem 3.9 to craft RLCC $C'$ with block length $n \in [\Omega(N/\log^{30} N), N]$, rate $1 - O(1/\log \log N)$, correcting radius $\delta = \Omega(1/\log^3 N)$, and query complexity $O(\log^{69} N/\log \log N)$. We now construct

$$C := \text{LTC} \pitchfork C'.$$

By construction, $C$ is explicit and has block length $N$. By Lemma 3.2 (in particular, Remark 3.3) on LTC and $C'$, the overall rate of $C$ is at least $R - (1 + n/N) \cdot O(1/\log \log N) = R - O(1/\log \log N)$. Applying Lemma 3.4 to LTC and $C'$, the correcting radius of $C$ is at least $\delta_{\text{LTC}}/2$ and the query complexity of $C$ is

$$O\left(\frac{qN}{\delta \kappa n}\right) + O\left(\frac{\log^{69} n}{\log \log n}\right) \leq O\left(\frac{q}{\kappa} \cdot \log^{33} N + \frac{\log^{69} n}{\log \log n}\right). \qquad \square$$

This corollary states that for any binary linear LTC, there is a binary linear RLCC with nearly the same rate and distance, and with the same (up to polylogarithmic factors) query complexity.

## 4 Final Construction

The main results of this paper follow from choosing an appropriate LTC for Corollary 3.10.

**Corollary 4.1** (Explicit RLCCs). *For any rate $R = 1 - \varepsilon \in (0, 1)$ and for infinitely many $n$, there is an explicit RLCC with block length $n$, rate $R - O(1/\log \log n)$, correcting radius $\Omega(\varepsilon^3)$, and query complexity*

$$O\left((1/\varepsilon)^{35} \log^{33} n + \frac{\log^{69} n}{\log \log n}\right).$$

*Proof.* We can instantiate Corollary 3.10 using the LTCs of Theorem 2.9. $\square$

**Corollary 4.2** (Gilbert–Varshamov bound RLCCs). *Let $H(\cdot)$ be the binary entropy function. For any $R, \delta, \varepsilon \in (0, 1)$ such that*

$$R + H(\delta) = 1 - \varepsilon$$

*and for infinitely many $n$, there exists a nonexplicit RLCC with block length $n$, rate $R - O(1/\log \log n)$, and distance at least $\delta$, with correcting radius $\delta/2$ and query complexity*

$$\text{poly}(1/\varepsilon) \cdot \log^{33} n + O\left(\frac{\log^{69} n}{\log \log n}\right).$$

*Proof.* Dinur, Evra, Livne, Lubotzky, and Mozes [DEL+22] construct explicit LTCs with rate arbitrarily close to 1, which implies the existence of infinitely many nonexplicit LTCs that approach the Gilbert–Varshamov bound (see [DEL+22, Corollary 1.2]). These LTCs can have any rate $R$ and distance $\delta$ such that $R + H(\delta) = 1 - \varepsilon$, in which case the testability is $\kappa \geq \text{poly}(\varepsilon)$ and the query complexity is $q_{\text{LTC}} \leq \text{poly}(1/\varepsilon)$. We can plug these parameters into Corollary 3.10 to yield RLCCs. Because the rate of the RLCC approaches the rate of the LTC, and the distance of the RLCC is at least the distance of the LTC, we can say that the RLCC also approaches the Gilbert–Varshamov bound. $\square$

# Acknowledgments

We would like to thank Dana Moshkovitz for valuable discussions and for feedback on a draft of this manuscript. We would also like to thank Siddhartha Jain for valuable advice towards the presentation of this work, Jeffrey Champion for proofreading, and Joshua Cook and Justin Oh for helpful conversations. Finally, we are grateful to anonymous reviewers for their input.

VMK is supported by NSF Grants CCF-2008076 and CCF-2312573, and a Simons Investigator Award (#409864, David Zuckerman). GM is supported by NSF Grant CCF-2200956, an NSF Graduate Research Fellowship (DGE-2137420), and a UT Austin Dean's Prestigious Fellowship Supplement.

This material is based upon work supported by the National Science Foundation Graduate Research Fellowship Program under Grant No. DGE-2137420. Any opinions, findings, and conclusions or recommendations expressed in this material are those of the authors and do not necessarily reflect the views of the National Science Foundation.

# A  Concrete Parameters for Locally Testable Codes

We instantiate the LTC construction from Theorem 2.9 with suitable parameters. In particular, for any $\varepsilon \in (0, 1)$, Dinur, Evra, Livne, Lubotzky, and Mozes [DEL+22] give an explicit construction for a family of LTCs with rate at least $1 - \varepsilon$ and distance, testability, and query complexity within a polynomial (or inverse polynomial) of $\varepsilon$, such that consecutive codes in the family differ in block size by a factor which is a polynomial of $1/\varepsilon$. Setting $\varepsilon = \Theta(1/\log N)$ yields the following family of LTCs:

**Theorem A.1** ([DEL+22, Theorem 1.1, Lemma 5.1, and Remark 5.3]). *For sufficiently large $N \in \mathbb{N}$, there exists an explicit odd prime power $p = \Theta(\log^{10} N)$ such that there is an infinite family of explicit binary linear locally testable codes $\{\mathsf{LTC}_1, \mathsf{LTC}_2, \dots\}$ where every $\mathsf{LTC}_j$ has*

- *block length $n_j = \Theta((p^{3j} - p^j) \cdot \log^{20} N)$,*
- *rate $1 - 1/(100 \log N)$,*
- *distance $\Omega(1/\log^3 N)$,*
- *testability $\Omega(1/\log^{15} N)$, and*
- *query complexity $O(\log^{20} N)$.*

**Corollary** (Corollary 3.7 restated). *For every sufficiently large $N \in \mathbb{N}$, there exists an integer $n \in [\Omega(N/\log^{30} N), N]$ and a family of LTCs $\{\mathsf{LTC}_1, \dots, \mathsf{LTC}_m\}$ such that*

- *Each $\mathsf{LTC}_j$ is binary, linear, and explicit.*
- *Each $\mathsf{LTC}_j$ has:*
  - *rate $R = 1 - \varepsilon_{\mathsf{LTC}} \geq 1 - O(1/\log N)$,*
  - *distance $\delta_{\mathsf{LTC}} \geq \Omega(1/\log^3 N)$,*
  - *testability $\kappa_{\mathsf{LTC}} \geq \Omega(1/\log^{15} N)$, and*
  - *query complexity $q_{\mathsf{LTC}} \leq O(\log^{20} N)$.*
- *Each $\mathsf{LTC}_j$ has block length $n_j$, such that $n_1 \leq O(\log^{50} N)$, $n_m = n$, and $\forall j.\ n_{j+1}/n_j = \Theta(\log^{30} N)$.*
- *The number of codes is $m = O(\log N / \log \log N)$.*

*Proof.* Let $p$ and $\{\mathsf{LTC}_1, \mathsf{LTC}_2, \dots\}$ be instantiated using Theorem A.1 with parameter $N$. Let $m$ be the smallest integer such that $n_m \leq N$, and define $n \coloneqq n_m$. Then, $\{\mathsf{LTC}_1, \dots, \mathsf{LTC}_m\}$ have the desired rate, distance, testability, and query complexity. In addition, $n_1 \leq O(p^3 \log^{20} N) = O(\log^{50} N)$. Next, for all $j \geq 1$,

$$\frac{n_{j+1}}{n_j} = \Theta\left(\frac{p^{3(j+1)} - p^{j+1}}{p^{3j} - p^j}\right) = \Theta(p^3) = \Theta(\log^{30} N).$$

Hence, $n \leq N \leq O(p^3 n)$, so $n \geq N/p^3$. Finally,

$$m \leq \frac{\log N}{O(\log(\log^{30} N))} = O\left(\frac{\log N}{\log \log N}\right).$$

□